\documentclass[
superscriptaddress,
nofootinbib,
twocolumn,
 amsmath,amssymb,
 aps,
pra,
notitlepage,
longbibliography
]{revtex4-1}

\usepackage{dcolumn}
\usepackage{bm}
\usepackage{epsfig}
\usepackage{graphicx}
\usepackage{latexsym}
\usepackage{amsfonts}
\usepackage{setspace}
\usepackage{graphicx}
\usepackage{amsmath}
\usepackage{verbatim}
\usepackage{color}
\usepackage{SIunits}
\usepackage{hyperref}
\usepackage{mathtools}
\usepackage{amsmath}
\usepackage{physics}

\newcommand{\be}{\begin{equation}}
\newcommand{\ee}{\end{equation}}
\newcommand{\ba}{\begin{array}}
\newcommand{\ea}{\end{array}}
\newcommand{\bqa}{\begin{eqnarray}}
\newcommand{\eqa}{\end{eqnarray}}

\begin{document}

\title{Continuous-variable graph states for quantum metrology}

\author{Yunkai Wang} 
\affiliation{Holonyak Micro and Nanotechnology Laboratory, University of Illinois at Urbana-Champaign, Urbana, IL 61801 USA}
\affiliation{Department of Physics, University of Illinois at Urbana-Champaign, Urbana, IL 61801 USA}
\affiliation{Illinois Quantum Information Science and Technology Center, University of Illinois at Urbana-Champaign, Urbana, IL 61801 USA}
\author{Kejie Fang} 
\email{kfang3@illinois.edu}
\affiliation{Holonyak Micro and Nanotechnology Laboratory, University of Illinois at Urbana-Champaign, Urbana, IL 61801 USA}
\affiliation{Illinois Quantum Information Science and Technology Center, University of Illinois at Urbana-Champaign, Urbana, IL 61801 USA}
\affiliation{Department of Electrical and Computer Engineering, University of Illinois at Urbana-Champaign, Urbana, IL 61801 USA}

\begin{abstract}
Graph states are a unique resource for quantum information processing, such as measurement-based quantum computation. Here, we theoretically investigate using continuous-variable graph states for single-parameter quantum metrology, including both phase and displacement sensing. We identified the optimal graph states for the two sensing modalities and showed that Heisenberg scaling of the accuracy for both phase and displacement sensing can be achieved with local homodyne measurements.

\end{abstract}

\maketitle

\section{Introduction}

Quantum metrology is among the most important applications of quantum information science, which aims at using quantum mechanical techniques and resources, such as noise squeezing and entanglement, to enhance the accuracy of parameter estimation beyond any classical means 
\cite{giovannetti2004quantum,giovannetti2006quantum,degen2017quantum,pirandola2018advances}.
One pronounced example is the use of squeezed light in interferometric gravitational wave detectors to probe weak mechanical displacement beyond the standard quantum limit \cite{abadie2011gravitational,aasi2013enhanced}. 

Besides quantum sensing using a single electromagnetic mode, which typically exploits the quantum correlation between its two quadratures, multi-mode quantum metrology that is relevant to distributed sensing uses non-classical inter-modal correlations to achieve advantages over classical resources, such as realization of the Heisenberg limit of the sensitivity in terms of, for example, the total number of modes. Multi-mode metrology based on both discrete variables \cite{holland1993interferometric,mitchell2004super,humphreys2013quantum}
and continuous variables (CV) \cite{anisimov2010quantum,zhuang2018distributed,matsubara2019optimal,guo2020distributed} has gained discussions in the literature. For the latter, many considered scenarios though rely on the output of highly squeezed states intertwined via a network of beam splitters as the probe.

Here, we consider another type of entangled states--graph states--as the probe for distributed quantum metrology. Graph states are deemed as a unique resource for measurement-based quantum computing. Recently, discrete-variable graph states have also drawn attention for applications in quantum sensing \cite{rosenkranz2009parameter,friis2017flexible,shettell2020graph}. In this paper, we study for the first time CV graph states for quantum metrology, including both phase and displacement sensing, of a single unknown parameter. We calculated the quantum Fisher information (QFI) for both cases with arbitrary graph states and identified the graph states to achieve the optimal scaling of sensitivity. Furthermore, we show that the QFI of both phase and displacement sensing can be attained (up to an $O(1)$ factor) using local homodyne measurements, highlighting the practicality of this scheme.

\begin{figure}[!htb]
\begin{center}
\includegraphics[width=0.8\columnwidth]{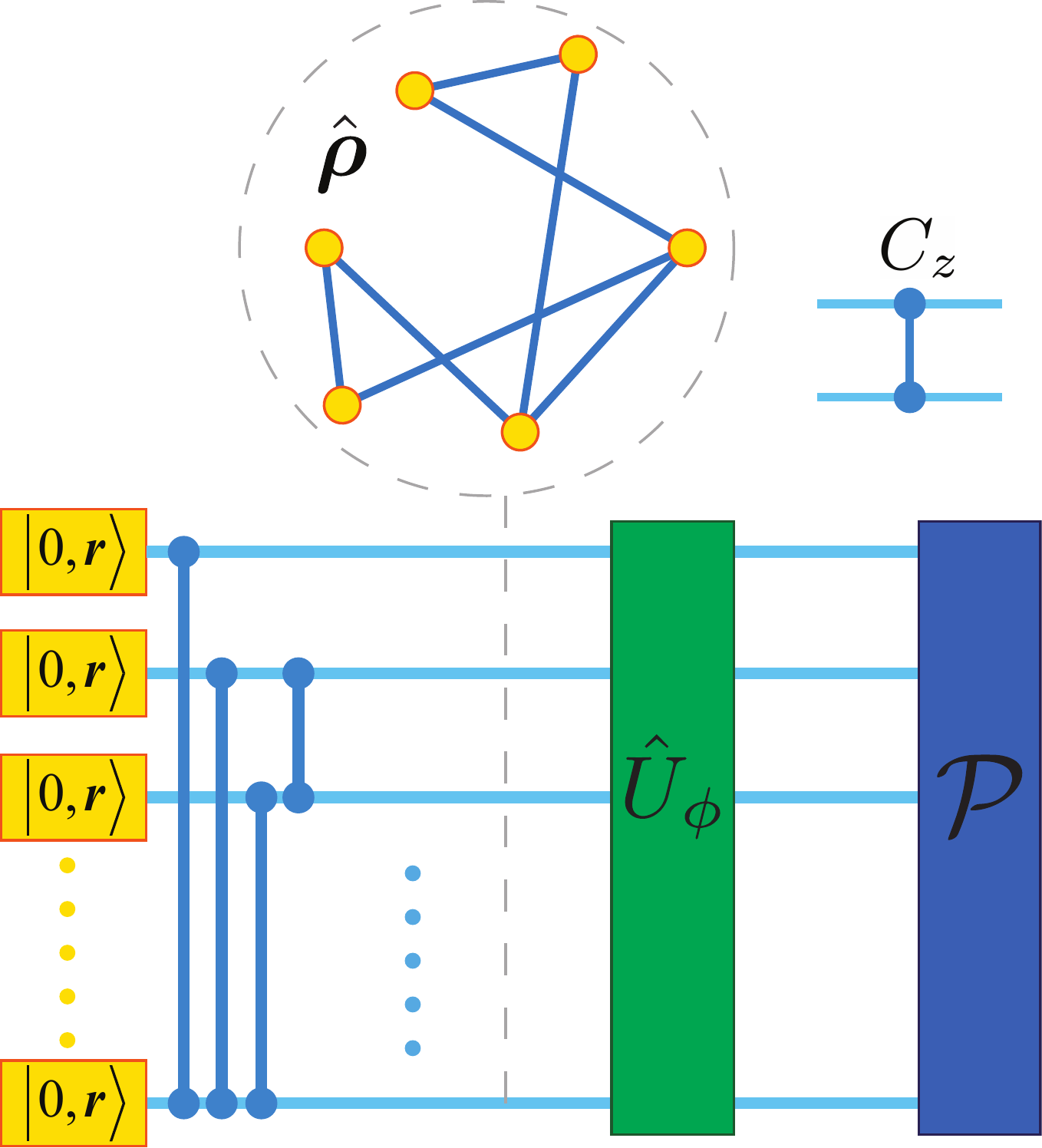}
\caption{Schematic plot illustrating generation of the CV graph state by applying $C_z$ gates onto squeezed vacuum states and its use in quantum metrology.} 
\label{general_setting}
\end{center}
\end{figure}

The general scenario of quantum metrology deploying CV graph states is schematically illustrated in Fig. \ref{general_setting}. The CV graph state with density matrix $\hat\rho$ is generated using the canonical method by applying $C_z$ gates onto a cluster of squeezed vacuum states $\ket{0, r}$, where $r$ is the squeeze factor. In this way, the graph state can be represented by a graph, where each vertex corresponds to a squeezed vacuum and an edge connecting two vertices represents the applied $C_z$ gate. The CV graph state then experiences a unitary transformation $\hat{U}_\phi$ which encodes the unknown variable $\phi$ onto the graph state, i.e., $\hat{\rho}_\phi=\hat{U}_\phi\hat{\rho}\hat{U}_\phi^\dagger$. Finally, proper measurements described by positive-operator valued measures (POVMs) $\mathcal{P}$ are performed on the transformed state $\hat{\rho}_\phi$ to extract the information about $\phi$. 


The accuracy of measuring $\phi$ is subject to measurement and fundamental noises. It is known that, for a given set of POVMs, the sensitivity $\delta\phi$ is constrained by the classical Cramer-Rao bound whose lower limit is given by the quantum Cramer-Rao bound--a quantity that is intrinsic to the probe state and the parameter-encoding transformation yet independent of POVMs, i.e.,
\begin{equation}
\delta\phi\geq1/\sqrt{I(\phi|\mathcal{P},\hat{\rho})}\geq1/\sqrt{F(\phi|\hat{\rho})},
\end{equation}
where $I(\phi|\mathcal{P},\hat{\rho})$ and $F(\phi|\hat{\rho})$ are the Fisher information (FI) associated with the POVM $\mathcal{P}$ and the QFI, respectively \cite{paris2009quantum,toth2014quantum}. In principle, for single-parameter metrology, POVMs can always be constructed such that the corresponding FI saturates the fundamental QFI \cite{paris2009quantum}. With proper quantum states as the probe, the quantum Cramer-Rao bound may achieve the Heisenberg scaling in terms of relevant resources, i.e., $\delta \phi\propto 1/n$, where $n$ could be, for example, the total number of photons or the number of modes. While with classical probe states, the best available sensitivity always complies with the standard quantum limit, i.e., $\delta \phi\propto 1/\sqrt{n}$.

Given $n$ bosonic modes with annihilation and creation operators $\{\hat a_k,\hat a_k^\dagger\}$, or equivalently quadrature operators $\hat q_k=(\hat a_k+\hat a_k^\dagger)/\sqrt{2}$,  $\hat p_k=(\hat a_k-\hat a_k^\dagger)/(i\sqrt{2})$, $k=1, ... ,n$, CV states created from these modes are typically described by the Wigner function of a Gaussian form
\begin{equation}
W(\mathbf{x})=\frac{\exp \left[-\frac{1}{2}(\mathbf{x}-\overline{\mathbf{x}})^{T} \mathbf{\Sigma}^{-1}(\mathbf{x}-\overline{\mathbf{x}})\right]}{(2 \pi)^{n} \sqrt{\operatorname{det} \mathbf{\Sigma}}},
\end{equation}
where $\hat{\mathbf{x}}:=\left(\hat{q}_{1}, \hat{p}_{1}, \ldots, \hat{q}_{n}, \hat{p}_{n}\right)^{T}$, $\overline{x}_i=\Tr(\hat{x}_i\hat\rho)$ , and $\Sigma_{i j}:=\frac{1}{2}\left\langle\left\{\hat{x}_{i}-\left\langle\hat{x}_{i}\right\rangle, \hat{x}_{j}-\left\langle\hat{x}_{j}\right\rangle\right\}\right\rangle$ is the covariance matrix ($\{~, ~\}$ is the anti-commutator) \cite{weedbrook2012gaussian}.
Our calculation uses the formulation of CV graph states generated from the finite squeezed vacuum \cite{simon1988gaussian,menicucci2011graphical}. Such CV graph states possess a covariance matrix
\begin{equation}
\Sigma=\frac{1}{2}SS^T,
\end{equation}
where $S$ is a symplectic matrix, and it can be further decomposed into the following form,
\begin{equation}\label{Eqn:sigma}
\Sigma=\frac{1}{2}\left[
\begin{matrix}
U^{-1} & U^{-1}V\\
VU^{-1} & U+VU^{-1}V
\end{matrix}\right],
\end{equation}
where $U$ is symmetric and positive definite and $V$ is symmetric.
For graph states created by the canonical method with $\hat C_z=\exp \left[\frac{i}{2}\hat{q}_1\otimes\hat{q}_2\right]$, where $\hat{q}_1$ and $\hat{q}_2$ are the position operators of the two involved modes, we have \cite{menicucci2011graphical}
\begin{equation}\label{Eqn:VU}
U=e^{-2r}I, \quad V=A, 
\end{equation}
where $r$ is the squeeze parameter of the input squeezed vacuum $\ket{0,r}\equiv\exp [\frac{r}{2}a^2-\frac{r}{2}a^{\dagger 2}]\ket{0}$, $I$ is the identity matrix and $A$ is the adjacency matrix of the underlying graph.

\section{Phase sensing}

\subsection{QFI for phase sensing}

We first consider the case of phase sensing, where the unknown parameter $\phi$ is encoded via the phase change of the probe state. The corresponding unitary transformation acting on the graph state is given by
\be\label{phaseU}
\hat{U}_\phi=\exp(-i\sum_{j=1}^n \phi f_j\hat{a}^\dagger_j\hat{a}_j),
\ee 
where $n$ is the number of modes constituting the graph state, $\hat{a}_{j}(\hat{a}^\dagger_{j})$ is the annihilation(creation) operator of the $j$-th mode, and $f_j$ is its responsivity. Eq. \ref{phaseU} is a passive, or energy-preserving, Gaussian unitary transformation, i.e., its generator $\hat{G}_\phi\equiv i\hat{U}^\dagger_\phi\frac{d\hat{U}_\phi}{d\phi}$ commutes with the number operator.  In general, the QFI for parameter estimation using Gaussian states and passive Gaussian unitary transformations can be calculated by \cite{matsubara2019optimal}
\begin{equation}\label{Eqn:QFI}
F(\phi|\hat{\rho})=\frac{1}{2}\Tr (G_\phi \Sigma^{-1}G_\phi\Sigma-G_\phi^2),
\end{equation}
where $G_\phi$ is the matrix form of the generator of the Gaussian transformation.

We straightforwardly calculated the QFI for phase sensing using a CV graph state with the covariance matrix $\Sigma$ given by Eqs. \ref{Eqn:sigma} and \ref{Eqn:VU}, which yields
\begin{widetext}
\begin{equation}\label{Eqn:QFIphase}
F(\phi|\hat{\rho})=2\sinh^2 2r\sum_{j=1}^n f_j^2+\sum_{j,k=1}^n(f_j^2+e^{4r}f_jf_k)A_{jk}A_{kj}+\frac{1}{2}e^{4r}\sum_{j,k=1}^nf_jf_k[A^2]_{jk}[A^2]_{kj},
\end{equation}
\end{widetext}
where $[A^2]_{ij}$ is the element of matrix $A^2$. The first term in Eq. \ref{Eqn:QFIphase} is the QFI for $n$ independent squeezed vacuum states, and the rest terms are resulted from the entanglement generated by $C_z$ gates. We also calculated the total mean photon number of the CV graph state (Appendix \ref{App:A}),
\begin{equation}\label{Eqn:N}
\bar{N}=n\sinh^2 r+\frac{1}{4}e^{2r}\Tr A^2.
\end{equation}
Similarly, the first term of Eq. \ref{Eqn:N} is the total mean photon number of $n$ uncoupled squeezed vacua and the second term is the addition due to $C_z$ gates. 

\subsection{CV graph states achieving the Heisenberg limit}

In this subsection, we identify CV graph states that achieve the Heisenberg limit for phase sensing, in terms of both total number of photons and number of modes. We first consider the simplified case where $f_j\equiv f$, $\forall j$. The QFI given in Eq. \ref{Eqn:QFIphase} reduces to
\be\label{Eqn:f}
F(\phi|\hat{\rho})=2nf^2\sinh^2 2r+(1+e^{4r})f^2\Tr A^2+\frac{1}{2}e^{4r}f^2\Tr A^4.
\ee
Because we want to explore the effect of entanglement of the graph state, it is assumed that the graph is sufficiently connected and the dominant terms in Eqs. \ref{Eqn:N} and \ref{Eqn:f} are the adjacency matrix-dependent terms. 

We introduce a characteristic figure that is intrinsic to the underlying graph of the CV graph state,
\be\label{chip}
\chi_{\textrm{p}}\equiv\frac{\Tr A^4}{\left(\Tr A^2\right)^2}.
\ee
Using this characteristic figure, the QFI for a sufficiently connected graph can be expressed as
\be
F(\phi|\hat{\rho})\approx 8f^2\chi_{\textrm{p}}\bar N^2\equiv8f^2\chi_{\textrm{p}}n^2\widetilde{N}^2,
\ee
where $\widetilde{N}=\bar{N}/n$ is the average number of photons per mode.
It is easy to show that 
\be
\chi_{\textrm{p}}\leq 1
\ee
and the equality is asymptotically saturated when the adjacency matrix $A$ has only one largest eigenvalue (in absolute value) and the ratio between it and the other eigenvalues approaches infinity as $n$ increases. In this case, the Heisenberg limit of phase sensitivity in terms of $\bar N$ is achieved. When $A$ has $m>1$ largest eigenvalues, with $m$ not to scale with $n$, whose mutual ratios are asymptotically nonzero constants, the Heisenberg limit can still be achieved, yet with the QFI reduced by a factor of $O(1)$ constant comparing to the previous case.

\begin{figure*}[!htb]
\begin{center}
\includegraphics[width=2\columnwidth]{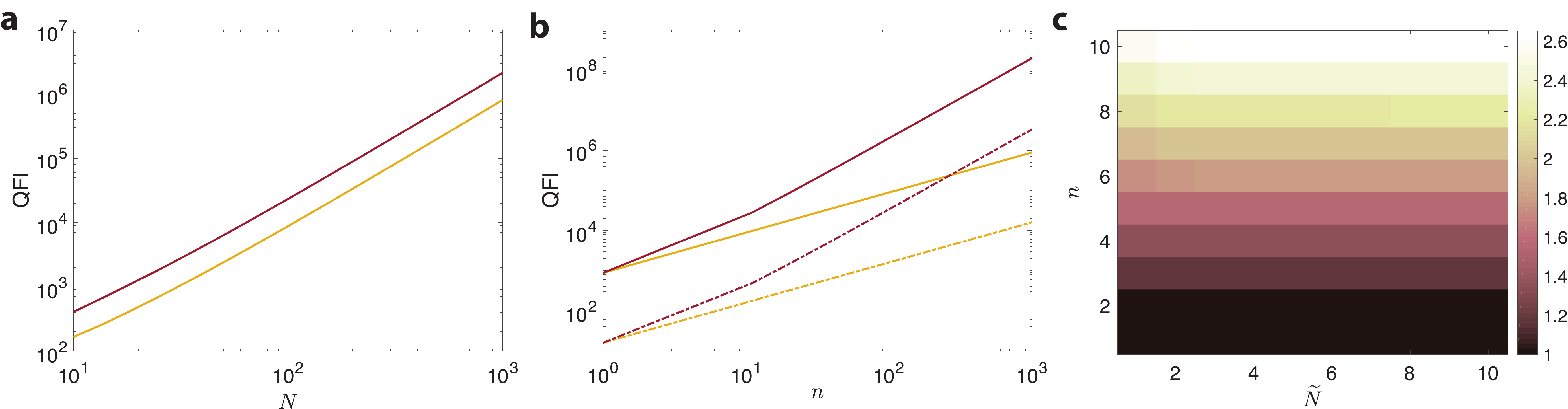}
\caption{Comparison of QFI scaling of phase sensing between the star graph state (maroon) and separable state (yellow) with respect to (a) $\bar N$ for fixed $n=10$ and (b) $n$ for fixed $\widetilde N=1$ (dash-dotted line) and $\widetilde N=10$ (solid line). (c) Ratio between QFI achieved by the star graph state and separable state for small $n$ and $\widetilde N$. We have chosen $f=1$ in these plots.}
\label{phase}
\end{center}
\end{figure*}

As an example, we show that the star graph state satisfies the condition above for achieving the Heisenberg limit for phase sensing. The star graph has only one vertex adjacent to all the other vertices which are otherwise unconnected. The adjacency matrix of the star graph is given by 
\be
A_{ij}=\delta_{i1}+\delta_{j1}-2\delta_{i1}\delta_{j1},
\ee
assuming the central vertex is $j=1$, which only has two nonzero eigenvalues $\lambda_{\pm}=\pm\sqrt{n-1}$, leading to $\chi_{\textrm{p}}=\frac{1}{2}$. To be more precise, \(F(\phi | \hat{\rho}) \approx \frac{16}{9} f^{2} \bar{N}^{2}= \frac{16}{9} f^{2} n^2\widetilde{N}^{2}\), after taking into account of the first term of Eq. \ref{Eqn:N}. Thus, the star graph state achieves the Heisenberg scaling in terms of total number of photons and the number of modes (for fixed number of photons per mode).  In contrast, the separable state with $A=0$ has $F(\phi|\hat{\rho})\approx\frac{1}{2}nf^2e^{4r} \approx 8f^2\bar{N}^2/n= 8f^2n\widetilde{N}^2$, which only achieves the standard quantum limit in terms of number of modes (for fixed number of photons per mode), while the Heisenberg scaling in $\bar N$ is weakened by a factor of $\sqrt{n}$. We plot the QFI of phase sensing (calculated using Eq. \ref{Eqn:f}) for the star graph state and separable state in Fig. \ref{phase}. The star graph state always outperforms the separable state. Another type of graph states that achieve the Heisenberg scaling in phase sensing is the $l-$multipartite graph states, which we detail in Appendix \ref{App:B}. 

Now we show that for the general case where $f_j$'s are not necessarily equal, the CV star graph state still achieves the Heisenberg limit. The QFI in this case is found to be
\be
\begin{aligned}
F(\phi|\hat{\rho})&\approx \frac{1}{2}e^{4r}\sum_{j=1}^n f_j^2 +\frac{1}{2}e^{4r}\left[(n-1)^2f_1^2+\left(\sum_{j=2}^n f_j\right)^2\right]\\
&> \frac{1}{2}e^{4r}(n-1)^2f_1^2\\
&\approx \frac{8}{9}f_1^2\bar{N}^2= \frac{8}{9}f_1^2n^2\widetilde{N}^2,
\end{aligned}
\ee
which indeed exhibits the same scaling as the equi-$f_i$ case.  In this sense, the CV star graph state is universally optimal for phase sensing. 

\subsection{Homodyne detection to saturate the quantum Cramer-Rao bound}

While the quantum Cramer-Rao bound sets the fundamental limit of parameter estimation, an important question is whether a POVM  exists whose FI saturates the QFI. Since we are only considering single-parameter estimation here, such POVM can always be found in theory \cite{paris2009quantum}. However, the mathematically constructed POVM is usually difficult to implement in most cases. Here, surprisingly, we numerically find that local homodyne detection actually saturates the QFI up to an $O(1)$ factor for the phase sensing using the star graph state. 

To show this, in general, the outcome of the homodyne detection (of the quadrature) of a Gaussian state is a Gaussian probability distribution regarding the unknown parameter $\phi$ \cite{adesso2014continuous,kay1993fundamentals}, which is captured by the first- and second-order moment $\omega$ and $\sigma_M$, respectively. The distinguishibility of probability distribution with different unknown parameter determines the sensitivity of $\phi$, which is quantified by FI. As a result, the FI can be calculated using the Gaussian probability distribution of the measurement outcome as \cite{kay1993fundamentals}
\begin{equation}\label{Eqn:FI_HD}
I=\frac{1}{2}\Tr\left[\left(\frac{\partial\sigma_{M}}{\partial \phi}\sigma_{M}^{-1}\right)^2\right]+\left(\frac{\partial \omega}{\partial \phi}\right)^{T}\sigma_M^{-1}\left(\frac{\partial \omega}{\partial \phi}\right).
\end{equation}
Following Refs. \cite{adesso2014continuous,kay1993fundamentals}, we find the first- and second-order moment of the homodyne detection for the phase sensing using CV graph states are given by $\omega=0$ and
\begin{widetext}
\begin{equation}
\begin{aligned}
\sigma_{M}=&\frac{1}{2}\big[(F_2G_1-G_2F_1)U^{-1}(F_2G_1-G_2F_1)
+(G_1G_2+F_1F_2)VU^{-1}(F_2G_1-G_2F_1)\\
&+(F_2G_1-G_2F_1)U^{-1}V(G_1G_2+F_1F_2)
+(G_1G_2+F_1F_2)(U+VU^{-1}V)(G_1G_2+F_1F_2)\big],
\end{aligned}
\end{equation}
\end{widetext}
respectively, where $U$ and $V$ are given by Eq. \ref{Eqn:VU}, and 
\begin{equation}\label{Eqn:FG}
\begin{aligned}
&[G_1]_{ij}=\delta_{ij}\cos f_j\phi,\\ 
&[F_1]_{ij}=\delta_{ij}\sin f_j\phi,\\ 
&[G_2]_{ij}=\delta_{ij}\cos \theta_j,\\ 
&[F_2]_{ij}=\delta_{ij}\sin \theta_j,
\end{aligned}
\end{equation}
with $\theta_j$ the phase difference between the local oscillator and the $j$-th mode.

\begin{figure}[!hbt]
\begin{center}
\includegraphics[width=0.8\columnwidth]{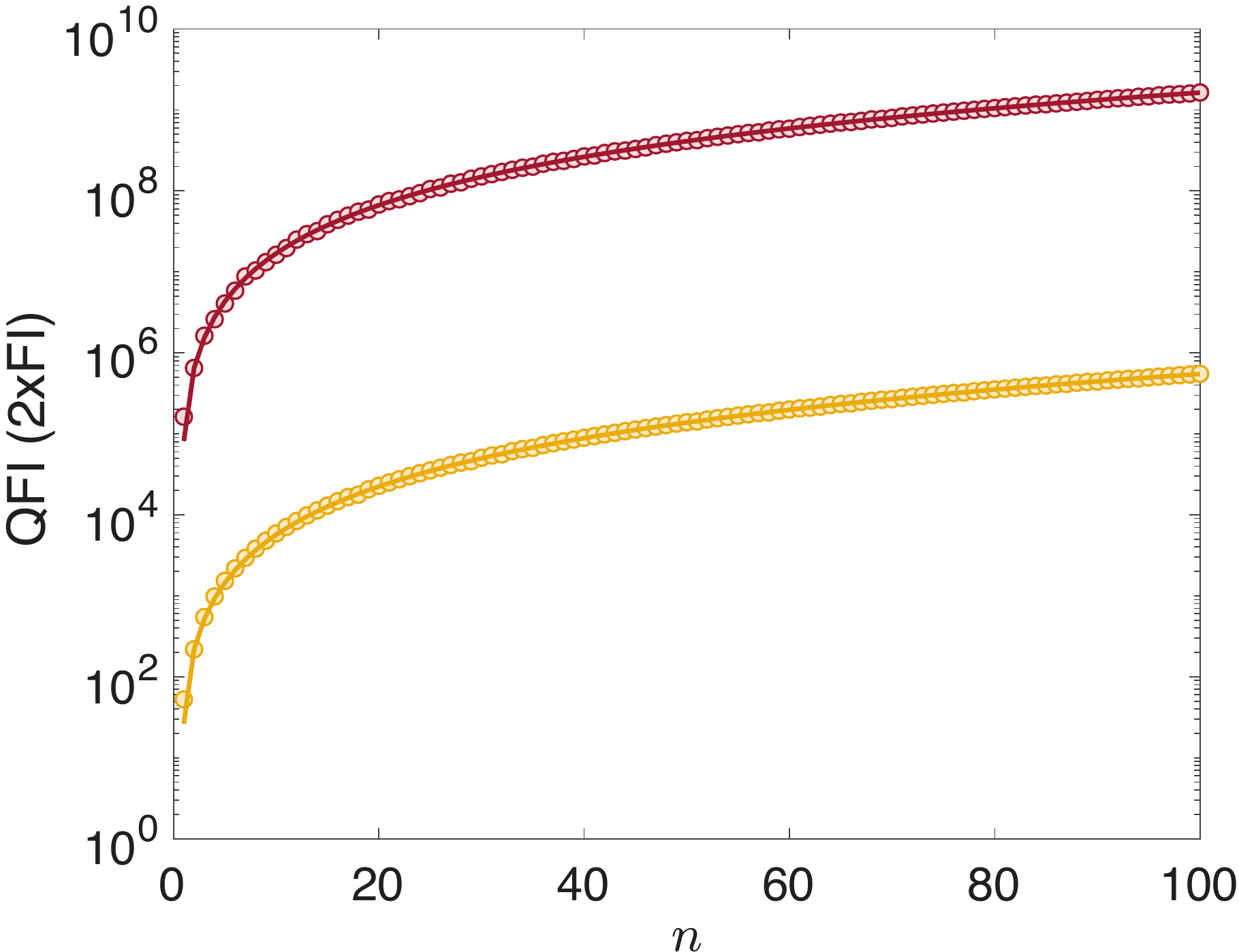}
\caption{Numerically calculated FI of phase sensing with the star graph state and the  homodyne detection with optimized local oscillator phases, for $r=1$ (yellow) and 3 (maroon). The dots are $2\times$FI. The solid line shows the analytical QFI. We have chosen $f=1$.}
\label{phase_FI}
\end{center}
\end{figure}

We numerically calculated and optimized FI, by varying $\theta_j$, for the star graph state with $f_j\equiv f$, $\forall j$. Because of the high-symmetric structure of the star graph, we expect that the maximum FI is achieved when 
\be\label{HD}
\theta_1=\alpha \mathrm{\ and\ } \theta_j=\beta, j\geq 2.
\ee
With this rational assumption, we find the numerically optimized FI saturates the analytical QFI given by Eq. \ref{Eqn:f} up to a constant, irrespective of $n$ (see Fig. \ref{phase_FI} for an example).

\section{Displacement sensing}
In this section, we consider another type of sensing modality that is displacement sensing. One canonical example of displacement sensing is found in the setting of an optomechanical cavity subject to a detuned pump, where the mechanical motion of the cavity mirror induces a displacement of the probe beam. For the displacement sensing, the unknown parameter $\phi$ is encoded in the probe state via an active, or photon-number non-conserving, unitary Gaussian transformation  $\hat{U}_\phi=e^{-i\phi\hat{X}_f}$, where $\hat{X}_f$ is a linear combination of mode quadratures, i.e., $\hat{X}_f=(f_1, f_2, \cdots, f_{2n})(\hat{q}_1,\,\hat{q}_2,\,\cdots,\,\hat{q}_n,\hat{p}_1,\,\hat{p}_2,\,\cdots,\,\hat{p}_n)^T$. The QFI for displacement sensing using CV graph state can be calculated using 
\begin{equation}
F(\phi|\hat{\rho})=4\sum_{i,j}f_if_j\Sigma_{ij},
\end{equation}
where $\Sigma$ is the covariance matrix given by Eq. \ref{Eqn:sigma}. 
For the graph state created by the canonical method, we have
\begin{widetext}
\begin{equation}\label{Eqn:QFIDG}
F(\phi|\hat{\rho})=\sum_{i=1}^n2e^{2r}f_i^2+\sum_{i=1}^{n}2e^{-2r}f_{i+n}^2+\sum_{i,j=1}^n 4e^{2r}f_if_{j+n}A_{ij} +\sum_{i,j=1}^{n}2e^{2r}f_{i+n}f_{j+n}[A^2]_{ij},
\end{equation}
\end{widetext}
where the first two terms are the QFI for $n$ independent squeezed vacua while the last two terms are attributed to the entanglement in the graph state.

\begin{figure*}[!htb]
\begin{center}
\includegraphics[width=2\columnwidth]{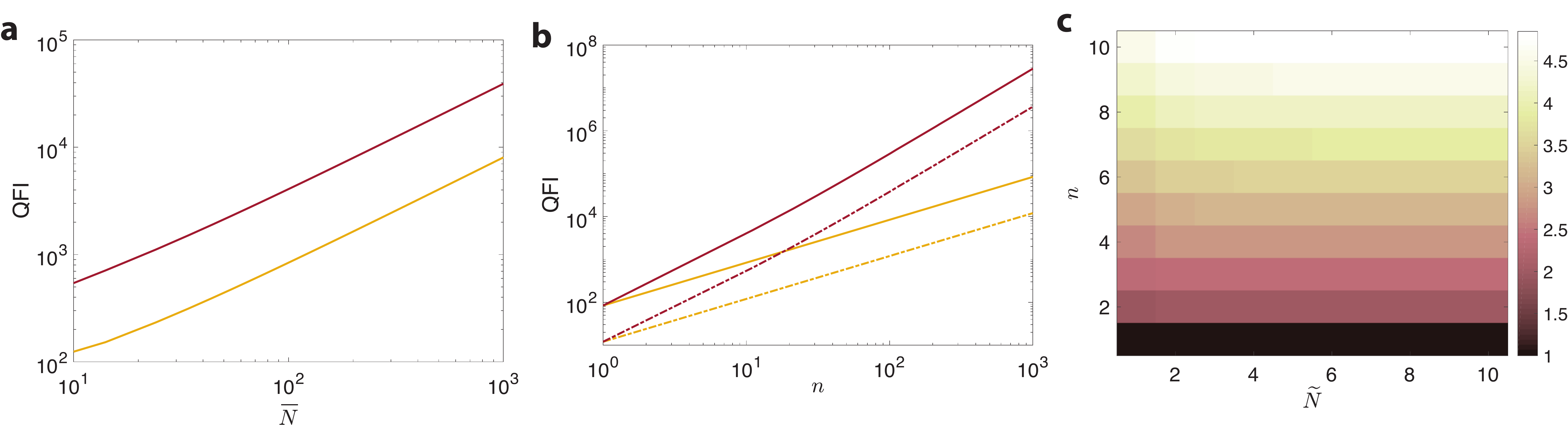}
\caption{Comparison of QFI scaling of displacement sensing between the star graph state (maroon) and separable state (yellow) with respect to (a) $\bar N$ for fixed $n=10$ and (b) $n$ for fixed $\widetilde N=1$ (dash-dotted line) and $\widetilde N=10$ (solid line).  (c) Ratio of QFI achieved by the star graph state and separable state for small $n$ and $\widetilde N$. We have chosen $f=1$.}
\label{displacement}
\end{center}
\end{figure*}

We first consider again the simplified case with $f_i\equiv f$ for all quadratures. In this case, Eq. \ref{Eqn:QFIDG} reduces to
\begin{widetext}
\be\label{Eqn:QFID}
F(\phi|\hat{\rho})=2nf^2e^{2r}+2nf^2e^{-2r}+\sum_{i,j=1}^n 4 f^2e^{2r}A_{ij}+\sum_{i,j=1}^n 2f^2e^{2r}[A^2]_{ij}.
\ee
\end{widetext}
We assume the leading terms in $F(\phi|\hat{\rho})$ and $\bar N$ are the adjacency matrix-dependent terms for a sufficiently connected graph. For the displacement sensing, we define another characteristic figure of a graph to be
\be
\chi_{\textrm d}\equiv\frac{\sum\limits_{i,j=1}^n[A^2]_{ij}}{\Tr A^2},
\ee
with which the QFI can be expressed as
\be\label{Eqn:QFIchid}
F(\phi|\hat{\rho})\approx 8f^2\chi_{\textrm d}\bar N\equiv8f^2\chi_{\textrm d}\,n\widetilde{ N}.
\ee

Since $[A^2]_{ij}$ is the number of common neighbors of the $i$-th and $j$-th vertices,  it is obvious that $[A^2]_{ij}\leq \min\{\mathrm{deg}(i),\mathrm{deg}(j)\}\leq \sqrt{\mathrm{deg}(i)\mathrm{deg}(j)}$, where $\mathrm{deg}(i)$ is the degree of the $i$-th vertex. This leads to
\begin{equation}\label{Eqn:QFINR}
\begin{aligned}
\chi_{\textrm d}&\leq \frac{\sum\limits_{i,j=1}^n \sqrt{\mathrm{deg}(i)\mathrm{deg}(j)}}{\mathrm{Tr}A^2}\\
&\leq \frac{n\left[\sum\limits_{i,j=1}^n \mathrm{deg}(i)\mathrm{deg}(j)\right]^{\frac{1}{2}}}
{\mathrm{Tr}A^2}\\
&= \frac{n\sum\limits_{i=1}^n \mathrm{deg}(i)}{\mathrm{Tr}A^2}\\
&= n,
\end{aligned}
\end{equation}
where we have used the Cauchy-Schwartz inequality and the fact that $\Tr A^2=\sum\limits_{i=1}^n \mathrm{deg}(i)$.
To asymptotically saturate the limit given by Eq. \ref{Eqn:QFINR}, or at least to realize the linear scaling in $n$ for $\chi_{\textrm d}$, we require $[A^2]_{ij}=\min\{\mathrm{deg}(i),\mathrm{deg}(j)\}=\sqrt{\mathrm{deg}(i)\mathrm{deg}(j)}=\textrm{const.}$ for most vertices in the graph, which means that the majority of vertices share the same neighbors. This indicates that the star graph state is again a good candidate. Another graph state satisfying this criterion is the $l-$multipartite graph (Appendix \ref{App:B}).

For the star graph state, the characteristic figure and QFI are $\chi_{\textrm d}=\frac{n}{2}$ and $F(\phi|\hat{\rho})\approx2n^2f^2e^{2r}\approx \frac{8}{3}f^2n\bar{N}=\frac{8}{3}f^2n^2\widetilde{N}$, respectively. Thus, the Heisenberg scaling over the number of modes (for fixed number of photons per mode) and the total number of photons is achieved. We stress that, in the displacement sensing, the Heisenberg scaling over total number of photons is linear in contrast to phase sensing. The separable state, on the other hand, has $F(\phi|\hat{\rho})\propto 2nf^2e^{2r}\approx 8f^2\bar N= 8f^2n\widetilde{N}$, which only achieves the standard quantum limit in terms of the number of modes.
We plot the QFI of phase sensing (calculated using Eq. \ref{Eqn:QFID}) for the star graph state and separable state in Fig. \ref{displacement}. The star graph state always outperforms the separable state.

We now consider the general case with unequal $f_j$ for the star graph state. The QFI to the leading order is given by,
\begin{widetext}
\begin{equation}
\begin{aligned}
F(\phi|\hat{\rho})&\approx 2e^{2r}\left[\left(\sum_{j=1}^{n}f_{j+n}\right)^2+f_{n+1}\left(nf_{n+1}-2\sum_{j=1}^{n}f_{j+n}\right)\right]\\
&\geq 2e^{2r}\left( n^2f_{\textrm{min}}^2+nf_{n+1}^2-2nf_{n+1}f_{\textrm{max}}\right),
\end{aligned}
\end{equation}
\end{widetext}
where $f_{\textrm{min(max)}}=\textrm{min(max)}\{f_{n+1}, \cdots, f_{2n}\}$ and we have labeled the center vertex of the graph to be $j=1$. Since $f_j$'s are independent of $n$ and $\bar N$, and assuming they do not vary significantly among different modes, we see that the Heisenberg scaling over the number of modes persists.

\begin{figure}[!tbh]
\begin{center}
\includegraphics[width=0.8\columnwidth]{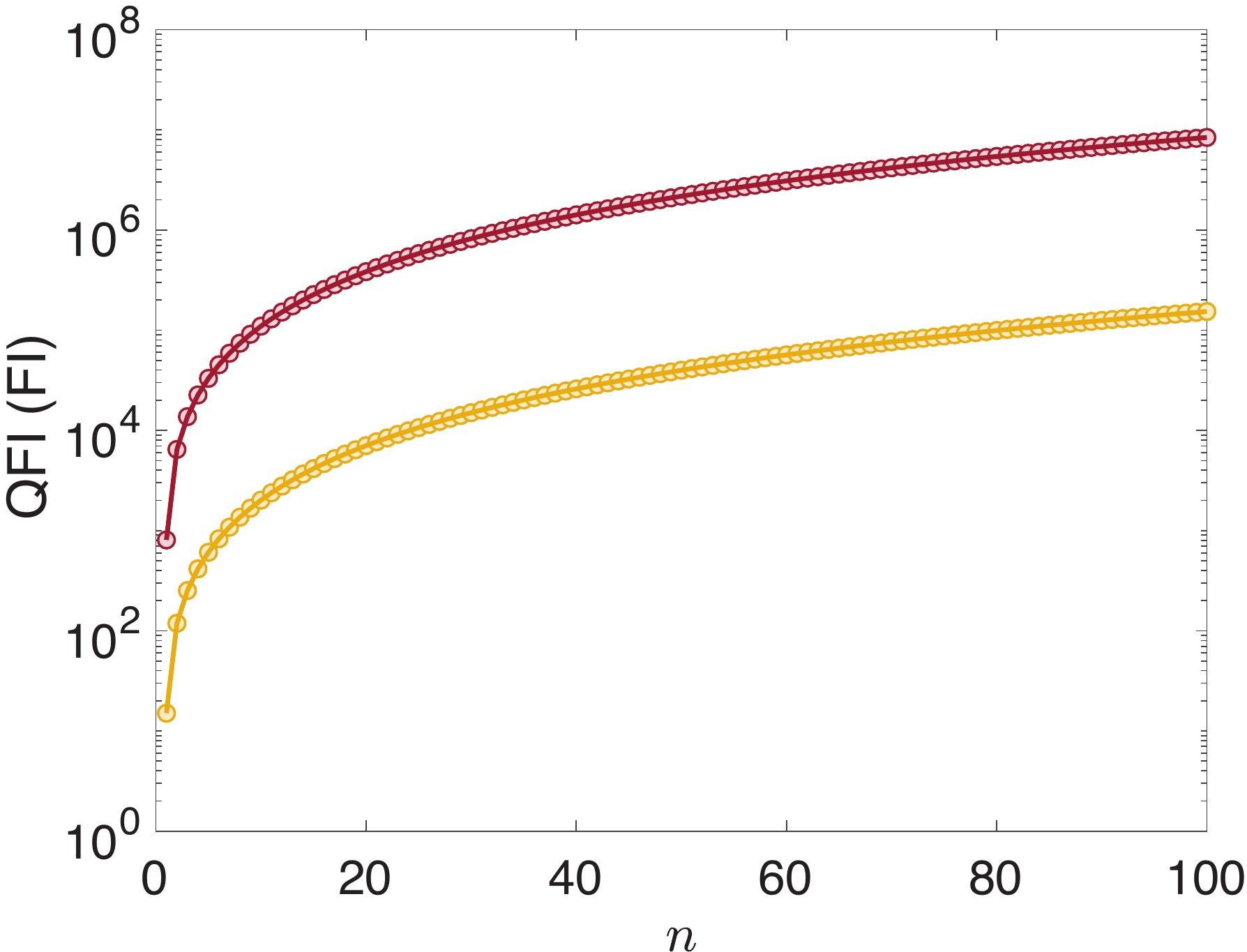}
\caption{Numerically calculated FI (dots) of displacement sensing with the star graph state and the homodyne detection with optimized local oscillator phases, for $r=1$ (yellow) and 3 (maroon). The solid line shows the analytical QFI. Here, we have chosen $f=1$.}
\label{displace_FI}
\end{center}
\end{figure}

Similar to the phase sensing, we investigate whether the QFI can be saturated by the local homodyne detection. For the displacement sensing, we find the first- and second-order moment of the probability distribution of the outcome of homodyne measurement is given by
\begin{equation}
\omega_i=\phi([F_2]_{ii}f_{n+i}-[G_2]_{ii}f_{i})
\end{equation}
and
\begin{widetext}
\begin{equation}
\sigma_{M}=\frac{1}{2}\big[F_2U^{-1}F_2+G_2VU^{-1}F_2+F_2U^{-1}VG_2+G_2UG_2+G_2VU^{-1}VG_2\big],
\end{equation}
\end{widetext}
respectively, where $F_2$ and $G_2$ are given by Eq. \ref{Eqn:FG}. The FI is then numerically calculated using Eq. \ref{Eqn:FI_HD} for the star graph state with $f_i\equiv f$. By varying the local oscillator phases $\alpha$ and $\beta$, as in the phase sensing, we find the optimized FI saturates the QFI (Fig. \ref{displace_FI}). This result suggests that local homodyne detection is able to achieve the optimal sensitivity predicted by the QFI for displacement sensing.

\section{Conclusion and Discussion}

In summary, we have investigated using CV graph states for quantum metrology, including both phase and displacement sensing that are of practical relevance. In particular, QFI of the two sensing modalities with general CV graph states is found, which can be benchmarked with characteristic factors intrinsic to the underlying graphs. With this, we identified the star graph states as an optimal resource which provides the Heisenberg scaling of the sensitivity over the number of modes for both phase and displacement sensing. Furthermore, we discovered that local homodyne detection is able to fulfill the QFI for phase and displacement sensing up to a constant factor, which makes this scheme based on CV graph states more appealing. 

Compared with multi-mode metrology schemes where the probe states are prepared as the outcome of a highly squeezed state propagating through a mesh of beam splitters \cite{zhuang2018distributed,guo2020distributed}, the scheme using graph states faces similar difficulty of generating large amount of squeezing, here for the implementation of the $C_z$ gate. However, several methods have been proposed to generate graph states, including those based on linear optics \cite{van2007building,su2007experimental}, frequency combs \cite{menicucci2007ultracompact,pysher2011parallel}, QND gates \cite{menicucci2010arbitrarily}, and photonic temporal modes \cite{menicucci2011temporal}. With proof-of-principle demonstrations of some of these proposals realized recently \cite{yukawa2008experimental,chen2014experimental,roslund2014wavelength, yokoyama2013ultra}, the CV graph states discussed in this work represent a viable resource for distributed quantum sensing.



\clearpage
\appendix

\section{Mean photon number of CV graph states} \label{App:A}

For a CV graph state with adjacency matrix $A$ created by the canonical method, its Wigner function is given by \cite{gu2009quantum}
\begin{equation}
\begin{aligned}
W(\vec{q},\vec{p})&=\frac{1}{\pi^n}\exp(-\sum_{j=1}^nq_j^2/s^2)\\
&\times\exp[-\sum_{j=1}^n s^2\left(p_j-\sum_{k=1}^nA_{jk}q_k\right)^2]
\end{aligned}
\end{equation}
where $s\equiv e^r$. The mean photon number could be calculated using
\begin{equation}
\bar{N}=\int\int d\vec{q}\,d\vec{p}\,W(\vec{q},\vec{p})\sum_{i=1}^n \frac{1}{2}(q_i^2+p_i^2-1),
\end{equation}
which yields
\begin{equation}
\bar{N}=n\sinh^2 r+\frac{1}{4}e^{2r}\Tr A^2.
\end{equation}

\section{Other examples of graph states}\label{App:B}

We calculate the QFI of some other graph states for phase and displacement sensing.

\emph{$l$-multipartite graph states}: The vertices of this type of graph states can be partitioned into $l$ sets and the vertices in each set are only connected with all vertices out of this set. Assuming each set has $m$ vertices, its adjacency matrix can be written as $A=A'\otimes G$, where $A'_{ij}=1-\delta_{ij}$, $1\leq i,j\leq l$, and $G_{kl}=1$, $1\leq k,l\leq m$. $A$ has two nonzero eigenvalues: $(l-1)m$ (1-fold) and $=-m$ ($(l-1)$-fold). 
As a result, we find $\chi_{\textrm{p}}=\frac{l-1}{l}\approx 1$ and  $\chi_{\textrm{d}}=m(l-1)\approx n$, which is similar to the star graph.

\emph{Rectangular graph states}: The corresponding graph resembles a belt formed by squares. Consider the $4\times m$ rectangular graph state with $n=4m$ modes. It has an adjacency matrix $A_{ij}=\delta_{i,j-1}+\delta_{i,j+1}+\delta_{i,j+4}+\delta_{i,j-4}$. In the large $n$ limit, $\Tr A^2=4n$, $\Tr A^4=36n- 84$, and $\sum\limits_{i,j=1}^n [A^2]_{ij}=16n$.
As a result, $\chi_{\textrm{p}}=\frac{36n-84}{16n^2}\approx \frac{9}{4n}$ and  $\chi_{\textrm{d}}=4$, leading to worse scaling than star and $l$-multipartite graph states.

\end{document}